\documentclass[aps,pra,twocolumn,superscriptaddress,a4paper]{revtex4-1}
\usepackage{graphicx}
\usepackage{natbib}
\usepackage{hyperref}
\usepackage{amsmath}
\usepackage{mathtools}
\usepackage{braket}

\begin{document}


\title{Coherence of a dynamically decoupled single neutral atom}
\author{Chang Hoong Chow}
\affiliation{Center for Quantum Technologies, 3 Science Drive 2, Singapore 117543}
\author{Boon Long Ng}
\affiliation{Center for Quantum Technologies, 3 Science Drive 2, Singapore 117543}
\author{Christian Kurtsiefer}
\affiliation{Center for Quantum Technologies, 3 Science Drive 2, Singapore 117543}
\affiliation{Department of Physics, National University of Singapore, 2 Science Drive 3, Singapore 117542}
\email[]{christian.kurtsiefer@gmail.com}
\date{\today}

\begin{abstract}
Long qubit coherence and efficient atom-photon coupling are essential for
advanced applications in quantum communication. One technique to maintain
coherence is dynamical decoupling, where a periodic sequence of refocusing
pulses is employed to reduce the interaction of the system with the
environment. We experimentally study the implementation of dynamical
decoupling on an optically-trapped, spin-polarized $^{87}$Rb atom. We use the
two magnetic-sensitive $5S_{1/2}$ Zeeman levels, $\ket{F=2,\ m_{F}=-2}$ and
$\ket{F=1,\ m_{F}=-1}$ as qubit states, motivated by the possibility to couple
$\ket{F=2,\ m_{F}=-2}$ to $5P_{3/2}$ the excited state $\ket{F'=3,\
  m'_{F}=-3}$ via a closed optical transition. 
With more refocusing pulses in the dynamical decoupling technique, we manage to extend the coherence time from 38(3)\,$\mu$s to more than two milliseconds.
We also observe a strong correlation between the motional states of the atom and the qubit coherence after the refocusing, which can be used as a measurement basis to resolve trapping parameters.
   \end{abstract}


\maketitle
\section{Introduction}
Quantum memories for efficient retrieval of a photonic qubit and long-lived storage are important building blocks for future applications of quantum communication~\cite{Childress2006, Kimble2008}.
Strong light-atom interaction is essential to accomplish a substantial information exchange between photons and atomic systems, or to implement an atom-mediated interaction between flying photonic qubits~\cite{Reiserer2014}.
One approach to realize such a quantum interface considers strong focusing of
the optical mode onto a confined atom~\cite{Volz2006, Tey2009, Vetsch2010, Alber2017, Bhaskar2017}.

In our experiment, we optically trap a single neutral atom at the focus of a
high numerical aperture lens for an incoming probe mode to achieve efficient light-atom coupling.
The clean energy level structure of a neutral atom and the trapping in ultra-high vacuum permits deriving the interaction strength with minimal assumptions.

In this work, we probe the lifetime of a coherent superposition of the $5S_{1/2}$ ground state Zeeman levels, $\ket{F=2,\ m_{F}=-2} \equiv \ket{\uparrow}$ and $\ket{F=1,\ m_{F}=-1} \equiv \ket{\downarrow}$ as our qubit states. The $\ket{\uparrow}$ state can be coupled to an auxiliary state $5P_{3/2}$, $\ket{F'=3,\ m'=-3}$ via a closed optical transition, opening a possible path to realize the sequential generation of an entangled photonic string~\cite{Lindner2009,Schwartz2016}. However, dephasing could lead to loss of information, reducing the fidelity of these entangled states. In comparison to other qubit configurations for neutral atoms, our interface based on the stretched states is more susceptible to noise such as magnetic field fluctuations. In earlier experiments, we have shown that a linearly polarized dipole trap can significantly reduce atomic motion-induced qubit dephasing without impacting the light-atom coupling \cite{steiner2019transmission}. One approach to further suppress decoherence is to apply dynamical decoupling (DD) techniques \cite{Biercuk2009, West2010, DeLange2010, Souza2011, Timoney2011, Lovric2013, Paz-Silva2016, Sukachev2017}.

We demonstrate that DD is efficient in mitigating the dephasing of the atomic
qubit. The experimental setup and the state readout procedure is described in
Sec.~\ref{sec:setup}. We first characterize our qubit system by performing
Rabi spectroscopy (Sec.~\ref{sec:rabi}), and carry on with applying various
types of DD (Sec.~\ref{sec:pdd}). From the result, we analyze the dephasing
mechanisms and tailor the refocusing sequence such that the
coherence is optimally preserved (Sec.~\ref{sec:udd}).
\begin{figure} 
\centering
  \includegraphics[width=\columnwidth]{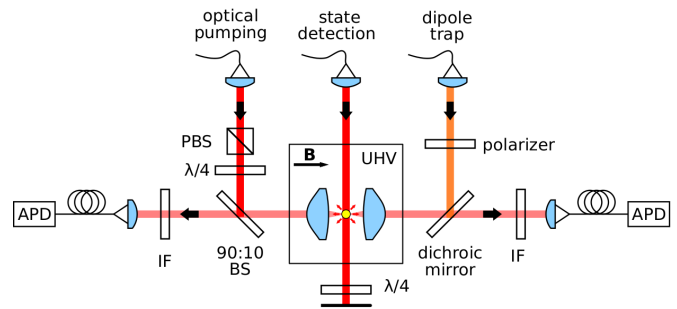}
  \caption{\label{fig:setup} Setup for probing light-atom interaction in free space. 
  APD:~avalanche photodetectors, UHV:~ultra-high vacuum chamber, 
  IF:~interference filter centered at 780\,nm, $\lambda$/2:~half-wave plate, $\lambda$/4:~quarter-wave plate, PBS:~polarizing beam splitter, BS:~beam splitter, B:~magnetic field.
}
\end{figure}

\section{Experimental setup}\label{sec:setup}
Our experiment starts with a single $^{87}$Rb atom trapped in a red-detuned
far off-resonant dipole trap (FORT) that is loaded from a magneto-optical trap
(MOT). This dipole trap is formed by a linearly polarized Gaussian laser beam
(wavelength 851\,nm) that is tightly focused by a pair of high numerical aperture lenses (NA = 0.75, focal length $f$ = 5.95\,mm) to a waist of $w_{0} = 1.4$\,$\mu$m \cite{steiner2019transmission,chin2017nonlinear}. Part of the atomic fluorescence is collected through the same lenses and coupled into single mode fibers that are connected to avalanche photodetectors (APD).

Once an atom is trapped, we apply 10\,ms of polarization gradient cooling to
reduce the atomic motion to a temperature of 14.7(2)\,$\mu$K \cite{chin2017PGC}. Then, a bias
magnetic field of 1.44\,mT is applied along the FORT laser propagation
direction to remove the degeneracy of the Zeeman states,
and the atom is optically pumped into $5S_{1/2}
\ket{F=2,\ m_{F}=-2} \equiv \ket{\uparrow}$. We implement a
lossless state-selective detection method \cite{Fuhrmanek2011,Gibbons2011} 
by sending light
resonant to the $5S_{1/2},\ F=2$ to $5P_{3/2},\ F'=3$ transition onto the atom
for 600\,$\mu$s and subsequent collection of the fluorescence light. The
atomic state can be inferred from the photodetection events recorded at the APDs.

The detection fidelity is characterized by first preparing the atom in a
particular state and then performing a state detection. When the atom is
prepared in the $\ket{\uparrow}$ state, the detectors record a mean of photon
number $n_{\uparrow}$ = 11.7(1). For atom in the $\ket{\downarrow}$
state, we expect the atom to scatter almost no photons due to the hyperfine
splitting of 6.8\,GHz. However, we find that in the experiment, the detectors occasionally register one or two events during the detection window (mean of photon number $n_{\downarrow}$ = 0.36(1)).

\begin{figure} 
\centering
  \includegraphics[width=\columnwidth]{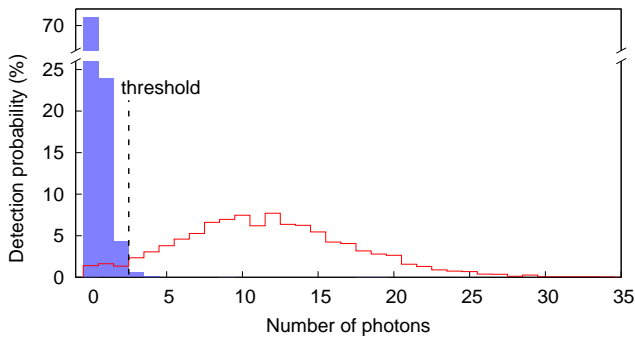} 
  \caption{\label{fig:state_detection} 
  Histogram of photon detection probability for atoms prepared in $F=1$, $\ket{\downarrow}$  (blue) and $F=2$, $\ket{\uparrow}$ (red) of the ground state manifold, respectively.
}
\end{figure}

We repeat this procedure for 2800 times. The histogram of $n_{\uparrow}$ and
$n_{\downarrow}$ is shown in Figure~\ref{fig:state_detection}. 
From this histogram, we can choose a threshold photon number $n_{th}$ that maximizes the discrimination between the two states.
Using $n_{th}$ = 3, the probabilities of a state assignment error are
$\xi_{\uparrow}$ = 4.4(4)\% and $\xi_{\downarrow}$ = 0.8(2)\,\% for atoms
prepared in states $\ket{\uparrow}$ and $\ket{\downarrow}$, respectively. 
With this, we achieve a detection fidelity of $F = 1 - (\xi_{\uparrow} + \xi_{\downarrow}) /2 = $97.4(2)\,\%.

\section{Rabi spectroscopy}\label{sec:rabi}

\begin{figure} 
\centering
  \includegraphics[width=\columnwidth]{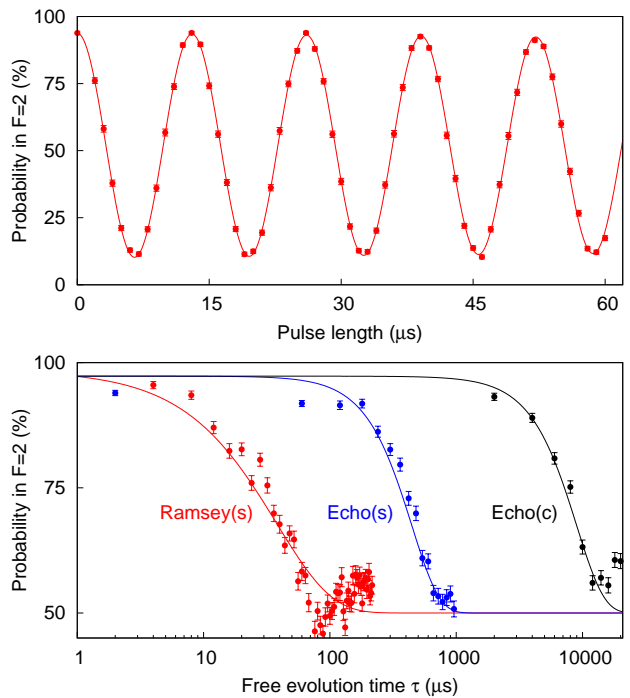}
  \caption{\label{fig:rabiramsey}
  Top: Rabi oscillation between $\ket{\uparrow}$ and $\ket{\downarrow}$. 
  Solid line is a fit to an exponentially decaying cosine function to extract the Rabi frequency, 
  $\Omega_{mw} = 2\pi \times 76.78(3)$\,kHz. 
  Bottom: Ramsey and spin-echo when the atom is initially prepared in $\ket{\uparrow}$ (s) or $\ket{F=2,\ m_{F}=0}$ (c). 
  We fit a decaying exponential to the Ramsey signal and a decaying Gaussian to the spin-echo signal to extract their respective $1/e$ time constants; $T^{\ast}_{2,s} = 38(3)$\,$\mu$s, $T_{2,s} = 480(21)$\,$\mu$s, and $T_{2,c} = 9.5(6)$\,ms. 
}
\end{figure}

Atoms in the $\ket{\uparrow}$ state are coupled to $\ket{\downarrow}$ by
applying a microwave field resonant to this transition. We then use this field
to drive Rabi oscillations and perform Ramsey and various dynamical decoupling
sequences to characterize the atomic coherence
\cite{Rosenfeld2011,Kuhr2003,Kuhr2005,Andersen2003,Andersen2004}. The Rabi
oscillation in Figure~\ref{fig:rabiramsey}\,(top), exhibits a Rabi frequency of
$\Omega_{mw} = 2\pi \times 76.78(3)$\,kHz at a visibility of 0.837(7). The
Rabi oscillation shows little decay within the first 60\,$\mu$s,
implying that the reduced visibility is most likely due to imperfections in the state preparation process.

To determine the dephasing time of the qubit system, we carry out a Ramsey
experiment where we apply two $\pi /2$-pulses ($t_{\pi /2} = \pi / (2\Omega)$)
with a free evolution time $\tau$ in between the two pulses to the atoms in
the $\ket{\uparrow}$ state. We repeat the experiment for different $\tau$ and
fit an exponential decay to the Ramsey contrast, which results in dephasing
time $T^{\ast}_{2} = 38(3)$\,$\mu$s [Fig.~\ref{fig:rabiramsey}\,(bottom)].

Next, we apply standard spin echo sequences~\cite{PhysRev.80.580, RevModPhys.76.1037}, which add an extra $\pi$-pulse ($t_{\pi} = \pi / \Omega$) in the middle of the free evolution window $\tau$. These sequences help to refocus the atomic state and reverse the inhomogeneous dephasing during the free evolution time, resulting in a much slower decay of the Rabi contrast. With these sequences, we obtain $T_{2} = 480(21)$\,$\mu$s for the stretched state of our qubits.

In order to compare the coherence in this qubit with other
systems~\cite{Kuhr2005,Yavuz2006,Jones2007}, we perform a spin echo on the
transition between magnetically insensitive Zeeman states, $5S_{1/2}$,
$\ket{F=1,\ m_{F}=0} \leftrightarrow \ket{F=2,\ m_{F}=0}$ of our qubits as
most of the other experiments were also probing this magnetically insensitive Zeeman state coherence.
  Using the same procedure, we find the coherence time of the magnetically insensitive qubit to be $T_{2,c} = 9.5(6)$\,ms, which is 20 times longer compared to the stretched state coherence [Fig.~\ref{fig:rabiramsey}]. 
  This observation is consistent with previous experiments with the superposition of magnetically insensitive Zeeman state in a red detuned dipole trap, which has a typical coherence time of 10\,ms.
  It has been shown that the coherence time can be improved to tens of milliseconds by reducing the trap depth~\cite{Kuhr2005,Jones2007}.  
  The coherence time on the order of hundreds of milliseconds has also been demonstrated by reducing the differential light shift with a magic-intensity trapping technique~\cite{Yang2016}.
  We suspect that the spatially varying dipole beam intensity gives rise to the differential light shift that limits our coherence time in the magnetic insensitive states.

\section{Periodic DD}\label{sec:pdd}

In the previous section, we showed that the spin-echo technique, as the simplest example of DD with one single $\pi$-pulse, can already improve the coherence time. To understand the effect of more complex DD on coherence, we adapt a semiclassical picture in the context of nuclear-magnetic-resonance (NMR) systems, which classifies decoherence processes into two classes: longitudinal energy relaxation and transverse dephasing. The dynamics of an undriven qubit in a decohering environment can be described by the Hamiltonian
\begin{equation}
\hat{H} = \hbar\beta_x(t) \hat{\sigma}_x + \hbar\big(\omega_0/2 + \beta_z(t) \big) \hat{\sigma}_z\,,
\end{equation}
with a qubit energy splitting $\hbar \omega_0$, and  transverse and
longitudinal random fields imparted by the environment characterized by $\beta_x$, $\beta_z$.

The longitudinal relaxation process, described by a characteristic energy relaxation
time $T_1$, is generally much slower than the transverse dephasing. 
Thus a superposition state evolves according to
\begin{equation}
\ket{\psi(t)} = e^{i\int_{0}^t\beta_z(t)dt}c_{\uparrow}\ket{\uparrow}+e^{-i\int_{0}^t\beta_z(t)dt}c_{\downarrow}\ket{\downarrow}
\end{equation}
in a rotating frame with $\beta_x=0$. The accumulation of random phases over
a duration $\tau$ decreases the state coherence $W(\tau) =
\overline{|\langle\sigma_y\rangle(\tau)|} = e^{-\chi(\tau)}$, represented by the length of a Bloch vector oriented along the $y$-axis after time $\tau$~\cite{Biercuk2011}.

\begin{figure} 
\centering
\includegraphics[width=\columnwidth]{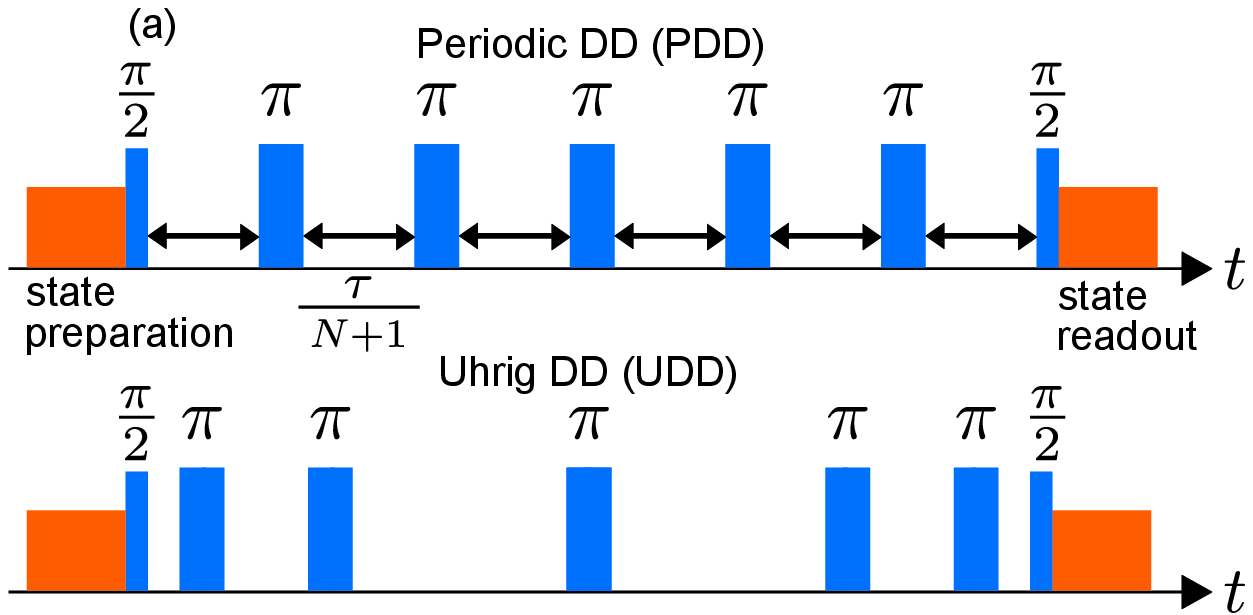}
\includegraphics[width=\columnwidth]{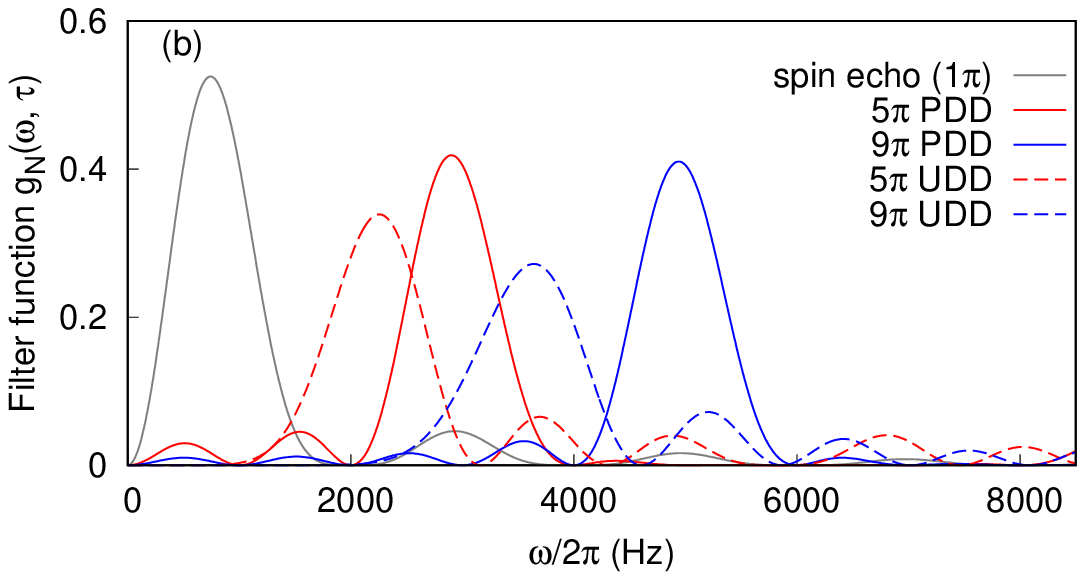}
  \caption{\label{fig:dd_general} (a): Schematic representation of various DD
    sequences. The qubit state is initialized in the $\ket{\uparrow}$
    state. We then bring the qubit state to the superposition state
    $(\ket{\uparrow}+i\ket{\downarrow})/\sqrt{2}$ with a $\pi /2$-pulse and
    let it evolve freely for a period $\tau$, with $\tau$ being partitioned
    into small windows using $\pi$-pulses. PDD partitions $\tau$ into uniform
    periods. UDD has its $j$-th $\pi$-pulse locating at $\delta_j \tau$ with
    $\delta_j = \text{sin}^2 [\pi j /(2N+2)]$. (b): Filter function
    $g_N(\omega, \tau)$ for different pulse sequences. Increasing the number
    $N$ of $\pi$-pulses shifts the peak to higher frequencies.
}
\end{figure}

Applying the control $\pi$-pulses flips the sign of the accumulated random
phases in different periods alternatively. To qualitatively understand the
efficiency of multipulse sequences on dephasing suppression, we focus on the
change in the state coherence integral 
\begin{equation}
\label{eqn:chi_tau}
\chi(\tau) = \frac{2}{\pi} \tau^2 \int_{0}^{\infty} S(\omega) g_N(\omega,\tau) d\omega\,,
\end{equation}
where $g_N(\omega,\tau)$ can be viewed as a frequency-domain filter function
of the random phases for a refocusing sequence consisting of $N$ $\pi$-pulses,
and $S(\omega)$ is the power spectral density of the noise term $\beta_z(t)$. 
Figure~\ref{fig:dd_general} illustrates the filter properties of function $g_N(\omega,\tau)$ for the Uhrig dynamical decoupling (UDD) sequence and periodic dynamical decoupling (PDD) sequence. 
For a fixed free evolution time $\tau$, the filter function's peak frequency shifts higher as $N$ increases, leading to a reduction of integrated low-frequency noise. 
The filter function gets narrower and is centered closer to $\omega = N\pi/\tau$ as $N$ increases.

\begin{figure} 
\centering
  \includegraphics[width=\columnwidth]{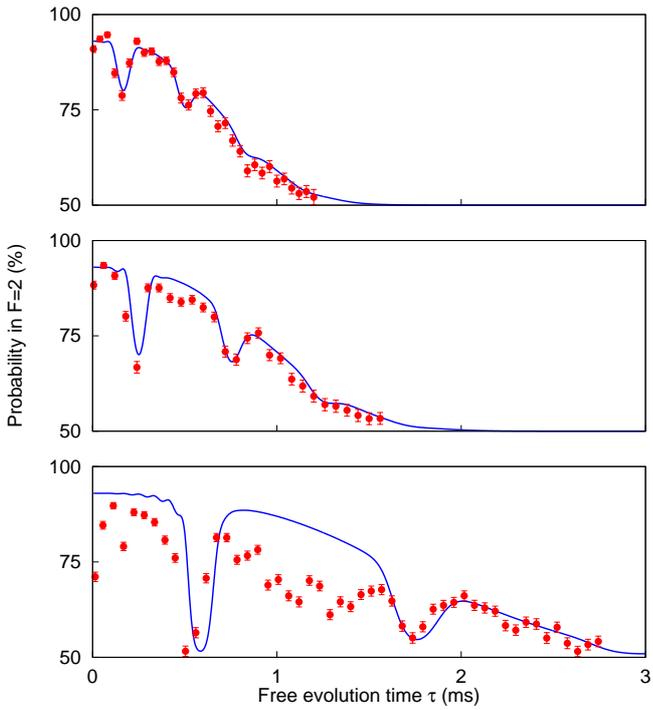}
  \caption{\label{fig:pdd_list} Coherence evolution under Periodic DD (PDD)
    for $N=3$ (top), $N=5$ (middle), and $N=13$ (bottom) $\pi$-pulses.
    Solid lines are numerical simulations using our heuristic noise model.
    Error bars represent standard error of binomial statistics accumulated from 300 repeated sequences.
}
\end{figure}

As a proof of concept, we first apply the PDD sequence, which has been shown to be able to compensate errors of non-ideal pulses and provide suppression of general decoherence \cite{Alvarez2010, Biercuk2011}.
Figure~\ref{fig:pdd_list} shows the coherence evolution of the qubit system under the PDD sequence. 
In contrast to a monotonic decaying profile, we observe that the decaying envelopes contain collapses which always occur at the same partition period $\tau/N$ for various $N$.
This can be explained by the atomic motion in the dipole trap; we discuss this
further in the next section.

\begin{figure} 
\centering
 \includegraphics[width=\columnwidth]{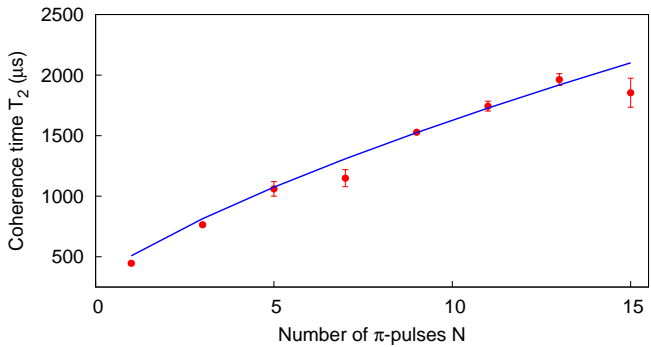}
  \caption{\label{fig:pdd_eff} Coherence time $T_2$ as a function of the
    number $N$ of $\pi$-pulses. The solid line shows the simulation result for a spectrum $S(\omega) \propto 1/\omega^\alpha$ with $\alpha = 1.73$.}
\end{figure}

To compare various decaying envelopes, we define the coherence time $T_2$ as
the time for the state coherence to decay by a factor of $1/e$.
This is consistent with the usual definition in a bare two-level system. 
Figure~\ref{fig:pdd_eff} shows the coherence time as a function
of number of $\pi$-pulses. The coherence time increases with the number $N$ of
$\pi$-pulses in a sequence. Our measurements suggest that the noise follows a
$1/\omega^{\alpha}$ spectrum with $\alpha > 0$. 
The dependence of $T_2$ on $N$ suggests that $T_2$ can potentially be further improved by using additional rephasing pulses.
A similar trend has been observed in other qubit systems, including single
silicon-vacancy centers \cite{Ryan2010}, single nitrogen-vacancy centers
\cite{Sukachev2017}, and single $^{43}$Ca$^+$ ion system \cite{Szwer2011}. 
In our system, we are currently limited to pulse sequences with $N \leq 20$ as the contrast of the coherence evolution drops as $N$ increases. 
In fact, the physical bound is $T_2 \leq 2 T_1$ with the energy-relaxation time $T_1$ determined to be on the order of a second in our system.

\begin{figure} 
\centering
  \includegraphics[width=\columnwidth]{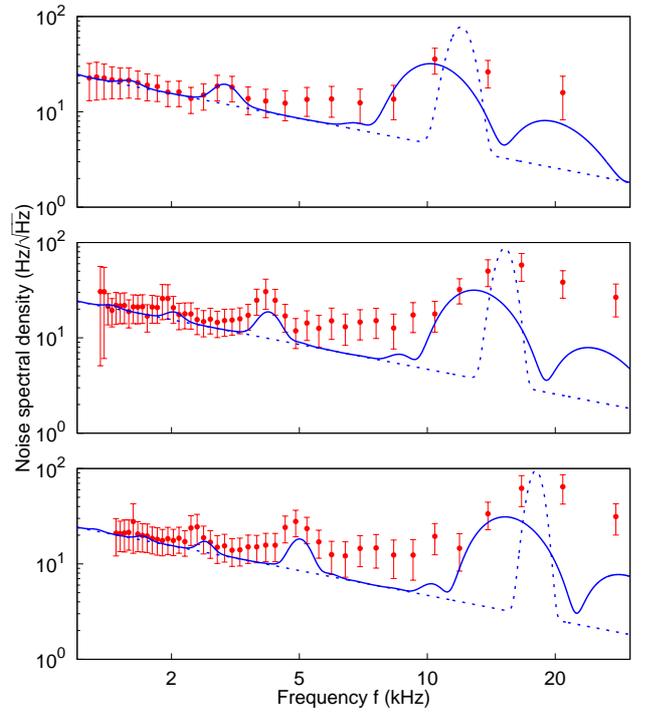}
  \caption{\label{fig:noise} Noise spectroscopy with DD adapted from atomic magnetometry. Red circle: noise spectral density reconstructed with experimental data. The recurring peaks are the feature of the filter function $g_N(\omega)$. Blue dashed line: noise spectrum of our heuristic noise model. Blue solid line: reconstructed noise spectral density in simulation. This is obtained by modulating the exact noise spectrum (blue dashed line) with the filter function of the chosen DD sequence. Trap depth is set to be 0.88\,mK (top), 1.04\,mK (middle) and 1.41\,mK (bottom), respectively. The trap frequencies used in simulation are 12.0\,kHz, 15.2\,kHz, and 18.0\,kHz, respectively.}
\end{figure}

To validate our findings, we simulate $\chi(\tau)$ under a simple noise model consisting of a $1/\omega^\alpha$ and a Gaussian centered at the axial trap frequency $\omega_0 = 2\pi \times 12.0$\,kHz.
The $1/\omega^\alpha$ spectrum represents the noise floor produced by ambient magnetic field fluctuations and power fluctuations of the dipole light field. The Gaussian spectrum represents the differential light shift due to the atomic motion in an inhomogenous dipole light field.
Our heuristic noise model is able to predict the recurring features as shown in Figure~\ref{fig:pdd_list}.
We further test our model by comparing the coherence time $T_2$ for different
numbers $N$ of $\pi$-pulses [Figure~\ref{fig:pdd_eff}]. 
Again, the model is in excellent agreement with the experimental data.

The qubit's sensitivity to the external magnetic field is 21\,GHz/T  at low fields. 
From the simulation we expect magnetic field fluctuations with a spectral
density on the order of $10^{-9}\,$T/$\sqrt{\text{Hz}}$ in the frequency range
of 1\,kHz to 30\,kHz. This is in agreement with the spectral density of
$\approx10^{-9}$\,T/$\sqrt{\text{Hz}}$ measured around 5\,cm away from the atom
with a fluxgate magnetometer (Stefan Mayer FLC100) in the band between 0 to 1\,kHz.

Dynamical decoupling is also implemented in the field of magnetometry to reconstruct the noise spectrum \cite{Hirose2012,Baumgart2016}. We manipulate the band-pass filter properties of $g_N(\omega,\tau)$ function to characterize the noise spectrum \cite{Bylander2011,Hernandez-Gomez2018}. Knowing that the filter function behaves as periodic sinc-shaped peaks at frequency $\omega_l \simeq (2 l + 1)\omega$ with $\omega \simeq N \pi / \tau$, we probe the noise spectral density by varying $N$ and $\tau$.

Figure~\ref{fig:noise} shows the noise spectra probed experimentally when the dipole beam power is being varied. The reconstruction of noise spectral density $S(\omega)$ follows~\cite{Bylander2011,Hernandez-Gomez2018}. 
The frequency range is determined by the choice of free evolution time $\tau$.
We observe the maximum noise density around $10.4$\,kHz, $16.7$\,kHz, and $20.8$\,kHz for dipole trap with trap depth of $0.88$\,mK, $1.04$\,mK, and $1.41$\,mK, respectively. 
As the dipole beam power increases, the maximum noise density shifts to higher frequencies.
The noise peaking at the axial trap frequency can be explained by the polarization gradients of a tightly focused FORT following \cite{Thompson2013}. Around the focal plane, the tight focusing of FORT results in a spatially varying vector light shift of the qubit states.
As the trap frequency along axial direction $\omega_z = \sqrt{2 U_0 / (m z_R^2)}$ increases along with the trap depth $U_0$, the light shift noise due to oscillatory atomic motion shifts to higher frequencies.

We also observe recurring peaks in the noise spectra at lower frequencies. These peaks are the feature of the filter function $g_N(\omega)$, determined by the DD sequence. 
We numerically construct the noise spectral density modulated by the filter function with our heuristic noise model and find that the simulation predicts the recurring features well.
By using the higher harmonics of the filter function, the trap frequency can be resolved with higher precision. We can use this as a basis for the precision measurement of trap parameters.

Another observation is that the width of the Gaussian noise in our model is
much narrower than the noise spectral density modulated with a filter function. 
This is because the bandwidth of the filter function is inversely proportional to $N$. 
In our experiment, the number of refocusing pulses $N$ used is less than 20, yielding a bandwidth that is comparable to the width of the Gaussian noise which we would like to resolve. 
It is possible to improve the resolution of the noise spectral density by increase the number of $\pi$-pulses $N$; however, there is a trade-off for increasing noise due to pulse errors.

Aside from the peak features, we notice that the background noise floor does not vary with dipole beam power.
We measure the intensity fluctuation of the dipole beam and find that it only corresponds to noise spectral density of $0.5$\,Hz/$\sqrt{\text{Hz}}$. This suggests that the background could be due to stray magnetic field fluctuation.

\section{DD benchmarking}\label{sec:udd}

\begin{figure} 
\centering
  \includegraphics[width=\columnwidth]{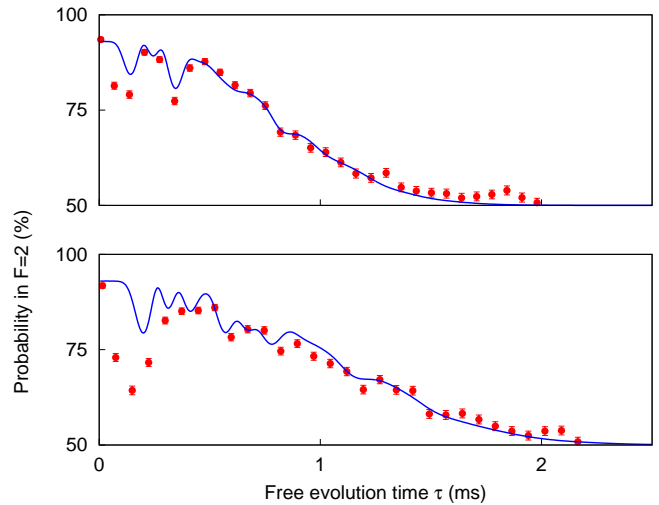}
  \caption{\label{fig:udd_list} Implementing Uhrig dynamic decoupling
    (UDD). Top: UDD with 3 $\pi$-pulses, $T_2 = 926\,\mu$s. Bottom: UDD with 5
    $\pi$-pulses, $T_2 = 1285\,\mu$s.
    Solid lines are numerical simulations using our heuristic noise model with
    the same parameters implemented in section~\ref{sec:pdd}.
    Error bars represent standard error of binomial statistics accumulated from 300 repeated sequences.
    }
\end{figure}

\begin{figure} 
\centering
 \includegraphics[width=\columnwidth]{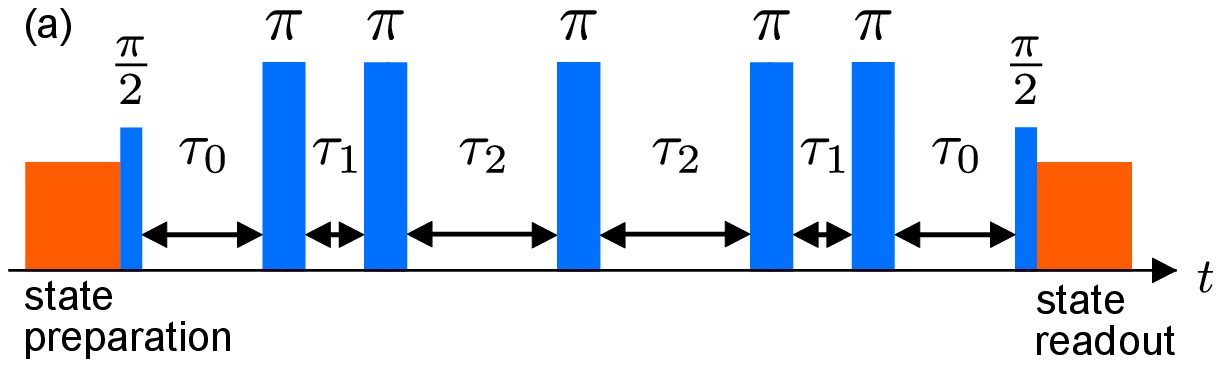}
  \includegraphics[width=\columnwidth]{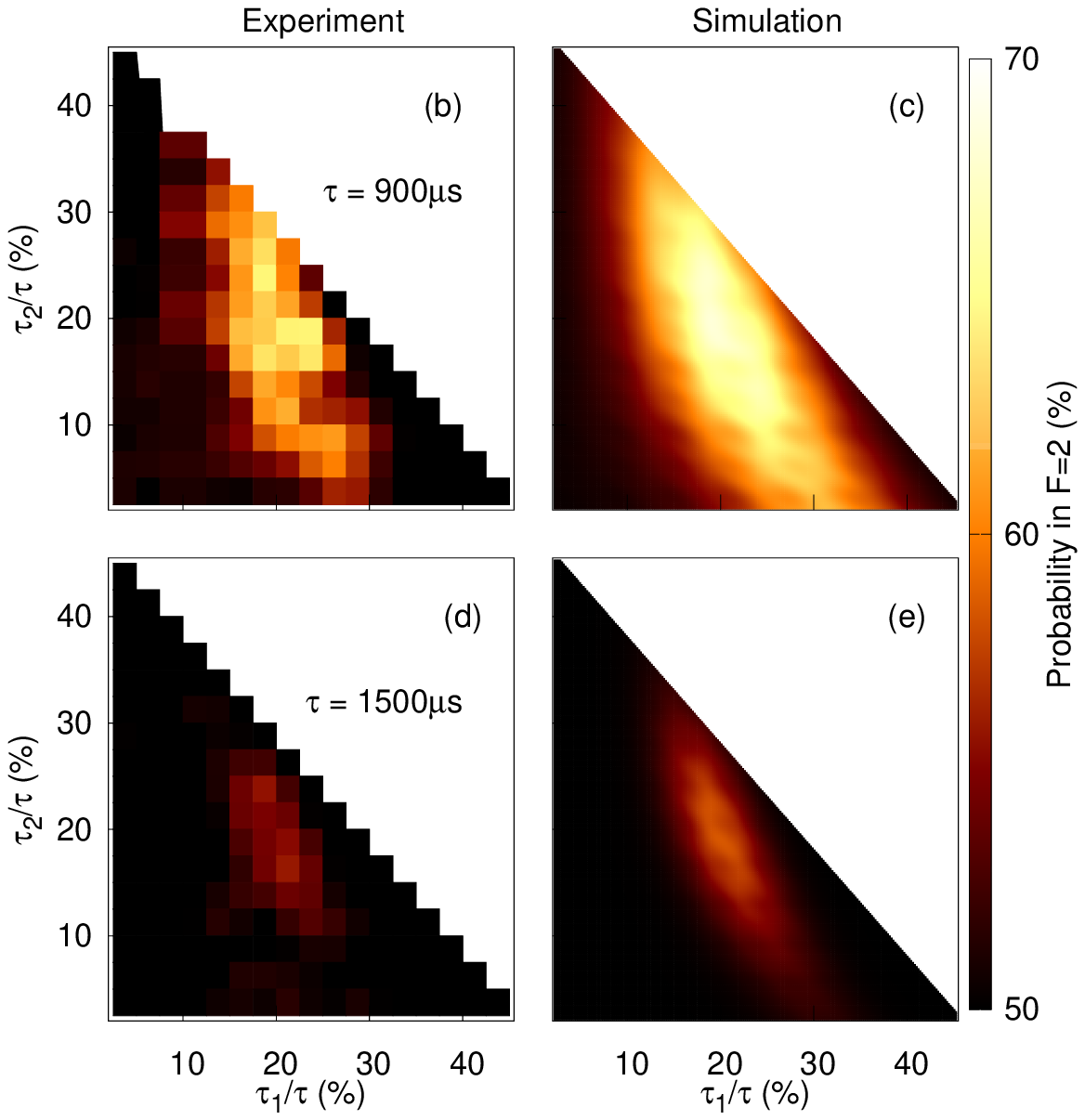}
  \caption{\label{fig:optimization} Optimization with five $\pi$-pulses for a fixed free evolution time $\tau=900\,\mu$s and $\tau=1500\,\mu$s. (a): Schematic representation of the DD sequence, satisfying $\tau_0+\tau_1+\tau_2=0.5 \tau$. (b-d): Population of $F=2$ state at the end of refocusing. For both $\tau=900\,\mu$s and $\tau=1500\,\mu$s, the maximum fidelity is not given by standard DD sequences such as UDD ($\tau_1/\tau=$18.3\,\%, $\tau_2/\tau=$25.0\,\%) or PDD ($\tau_1/\tau=$16.7\,\%, $\tau_2/\tau=$16.7\,\%), the maximal point locates at $\tau_1/\tau=$19.2\,\%, $\tau_2/\tau=$19.6\,\% in the simulation.
}
\end{figure}

We also apply Uhrig DD (UDD) protocols \cite{Uhrig2007} to suppress dephasing in our qubit system.
The UDD sequence has been analytically shown to provide strong suppression of phase accumulation when the noise environment contains a high-frequency component and a sharp high-frequency cutoff.
The $\pi$-pulse sequence and the characteristics of the filter function $g_N(\omega,\tau)$ for UDD are shown in Figure~\ref{fig:dd_general}.
A feature of UDD is the lack of higher harmonics but more side lobes. Compared with the PDD protocol having the same number of $\pi$-pulses $N$, UDD produces a pass band with a larger width peaking at a lower frequency.
This indicates that UDD could perform worse under a broadband noise spectrum.

Figure~\ref{fig:udd_list} shows the UDD coherence evolution of a single atom qubit.
Again, the simulation with our heuristic noise model introduced in Sec.~\ref{sec:pdd} predicts the wiggles qualitatively in the $\ket{\uparrow}$ population as the total free evolution time $\tau$ varies.
However, the simulation falls short in predicting the magnitude of the wiggles.
This is most likely due to the simplified formulation for the filter function $g_N(\omega,\tau)$ that assumes an instantaneous $\pi$-pulse.

We also look at the $1/e$ coherence time under the UDD protocol for a free evolution time $\tau$ larger than 500\,$\mu$s to minimize the influence from the wiggles. 
We observe a coherence time of 926\,$\mu$s and 1285\,$\mu$s for $N=3$ and $N=5$ $\pi$-pulses, respectively.
Compared with the coherence time obtained using PDD with the same number of $\pi$-pulses (764\,$\mu$s for $N=3$ and 1060\,$\mu$s for $N$=5), we observe an improvement of 21.2\,\% on the coherence time, consistent for both $N = 3$ and $N = 5$.

For most applications in quantum information processing, we aim to preserve coherence maximally for a given duration.
We demonstrate the optimization protocol with $N=5$ $\pi$-pulses. As shown in Figure~\ref{fig:optimization}\,(a), we impose a fixed free evolution time $\tau$ and reflection symmetry as constraints to reduce the number of free parameters from 6 to 2.
To better understand the effect of the noise on the qubit coherence, we numerically calculate the dynamics of the qubit state 
using our heuristic noise model introduced in previous sections, following Eqn.~(\ref{eqn:chi_tau}).

We find a good agreement between the observed coherence and the model for the same parameters used in the previous section. The maximum coherence is obtained with the protocol that follows $(\frac{\tau_0}{\tau}, \frac{\tau_1}{\tau}, \frac{\tau_2}{\tau})=(11.2\,\%, 19.2\,\%, 19.6\,\%)$.  This optimal sequence matches well with the Carr-Purcell-Meiboom-Gill (CPMG) sequence, which is widely used in the field of NMR and is constructed with the first and last precession periods are half of the duration of the interpulse period, e.g. $(\frac{\tau_0}{\tau}, \frac{\tau_1}{\tau}, \frac{\tau_2}{\tau})=(10\,\%, 20\,\%, 20\,\%)$ \cite{Slichter_1990}. This is in agreement with a previous study that shows that the CPMG sequence performs well for noise with a soft cut-off~\cite{pasini2010}.
\

\section{Conclusion}\label{sec:con}

We have presented a detailed experimental study of the
implementation of dynamical decoupling (DD) in a single neutral atom qubit
system. In addition to the performance comparison between two standard DD
protocols, periodic DD and Uhrig DD, we find an improvement in the coherence time $T_2$ by two orders of magnitude from $T_2^\ast$. The observed coherence time of $2$\,ms is sufficient to facilitate the high-fidelity transfer of quantum states between quantum repeater nodes separated by hundreds of kilometers~\cite{Childress2006}. 
In particular, we characterized the noise spectrum of an optically trapped Rubidium atom. We demonstrated that the CPMG sequence performs the best in the longer timescale. 

Future experiments will explore lowering the noise floor and motion-dependent
dephasing. Improvements will extend the coherence times and hence open up new
possibilities for the implementation of more robust free-space neutral atom
quantum memories for future quantum repeater
networks~\cite{Razavi2009}. A better understanding of the qubit response to
noise may also help to develop a broadband single-atom sensor which would
allow to image magnetic fields with a spatial resolution at atomic length
scales.

\
\
\begin{acknowledgments}
We thank Y.S.\,Chin and M.\,Steiner for contributions in an early stage of the
experiment. This work was supported by the Ministry of Education in Singapore.
\end{acknowledgments}
\bibliographystyle{apsrev4-1}
%

\end{document}